\documentclass[aps,prl,twocolumn,superscriptaddress,showpacs,floatfix]{revtex4}

\usepackage{amsmath,amssymb}
\usepackage{graphics,epsfig}

\makeatletter

\newcommand{\bfn}{{\mathbf{n}}}

\newcommand{\bfr}{{\mathbf{r}}}
\newcommand{\bfx}{{\mathbf{x}}}

\newcommand{\bfz}{{\mathbf{z}}}

\newcommand{\bfsigma}{{\boldsymbol{\sigma}}}

\newcommand{\varG}{{\mathcal{G}}}
\newcommand{\varH}{{\mathcal{H}}}
\newcommand{\varO}{{\mathcal{O}}}

\newcommand{\ux}        {{\hat\bfx}}

\newcommand{\uz}        {{\hat\bfz}}

\newcommand{\ur}        {{\hat\bfr}}

\newcommand{\meter}     {{\rm m}}
\newcommand{\nmeter}    {{\rm nm}}
\newcommand{\evolt}     {{\rm eV}}
\newcommand{\mevolt}    {{\rm meV}}
\newcommand{\mtesla}    {{\rm mT}}
\newcommand{\kelvin}    {{\rm K}}

\newcommand{\im}{\mathop{\operator@font Im}}
\newcommand{\rcp}[1]{\frac{1}{#1}}

\newcommand{\pde}[3][]{\frac{\partial^{#1}{#2}}{\partial{#3}^{#1}}}
\newcommand{\Pde}[2][]{\frac{\partial^{#1}}{\partial{#2}^{#1}}}

\newcommand{\eqnref}[1]{Eq.~(\ref{#1})}

\newcommand{\figref}[1]{Fig.~\ref{#1}}
\newcommand{\Figref}[1]{Figure~\ref{#1}}

\makeatother

\begin{document}
\title{Spin filter using a semiconductor quantum ring side-coupled to
  a quantum wire}
\author{Minchul Lee}
\affiliation{Department of Physics and Astronomy, University of Basel,
CH-4056 Basel, Switzerland}
\affiliation{Department of Physics, Korea University, Seoul 136-701, Korea}
\author{C. Bruder}
\affiliation{Department of Physics and Astronomy, University of Basel,
CH-4056 Basel, Switzerland}

\begin{abstract}
We introduce a new spin filter based on spin-resolved Fano resonances
due to spin-split levels in a quantum ring (QR) side-coupled to a
quantum wire (QW). Spin-orbit coupling inside the QR, together with
external magnetic fields, induces spin splitting, and the Fano
resonances due to the spin-split levels result in perfect or
considerable suppression of the transport of either spin
direction. Using the numerical renormalization group method, we find
that the Coulomb interaction in the QR enhances the spin filter
operation by widening the separation between dips in conductances for
different spins and by allowing perfect blocking for one spin
direction and perfect transmission for the other. The spin-filter
effect persists as long as the temperature is less than the broadening
of QR levels due to the QW-QR coupling. We discuss realistic
conditions for the QR-based spin filter and its advantages to other
similar devices.
\end{abstract}

 \maketitle

\textit{Introduction.}--- Spintronics \cite{Spintronics} that utilizes
the electron's spin degree of freedom rather than its charge for
information processing and storage has been a subject of intense
interest in recent decades.  The practical realization of spin-based
electronic circuits requires the development of efficient means to generate
spin-polarized currents, and to manipulate and detect spins.
\textit{Spin filters} that block the transport of one spin
direction have been proposed as a device to generate and detect spin currents
\cite{SpinFilters}. The basic scheme of a spin filter exploits
spin-dependent transport through systems lacking time-reversal
symmetry or having nontrivial geometric structures with spin-dependent
interactions; such systems include ferromagnetic junctions
\cite{SpinFilters}, and nanostructures like quantum dots
\cite{SpinFilterQD} (QD's) and rings
\cite{SpinFilterQR1,SpinFilterQR2}.

A simple but effective spin-filter implementation without
coupling to magnetic materials has been suggested to exploit the
\textit{spin-dependent resonance} through a QD with Zeeman
splitting that is embedded \cite{SpinFilterQD} in, or side-coupled
\cite{Torio04,Aligia04} to a quantum wire (QW). While in both cases
the spin-dependent transport is based on scattering from spin-split QD
levels and can be tuned by varying the gate voltage or the external
magnetic field, the side-coupled configuration is more effective because the
Fano resonance in this case can lead to \textit{perfect blocking} of
one spin direction and almost total transmission of the other.
Side-coupled QD systems show two dips corresponding to the total
suppression in the conductance of spin up and down
\cite{Torio04,Aligia04}. Since such spin filtering deteriorates
rapidly with increasing temperature $T$, however, an ideal operation of
the device requires large magnetic fields $B$ or high $g$-factors such
that $g\mu_B B \gg k_BT, k_BT_K$ where $\mu_B$ is the Bohr magneton
and $T_K$ is the Kondo temperature \cite{Aligia04}.

Recently, \textit{quantum rings} (QR's) with Rashba spin-orbit
coupling have been proposed for spin injection devices
\cite{SpinFilterQR2}. Spin precession due to the momentum-dependent
effective magnetic field and the following spin interference of two
quantum states propagating in opposite directions can not only
modulate the charge conductance \cite{SOQuantumRing2} but also induce
spin currents through leads attached to the ring \cite{SpinFilterQR2}.
When an unpolarized charge current is injected through one of leads,
quantum interference can produce pure spin currents through one of
other leads. However, this interference-based spin-filter operation
requires more than two leads to be linked to different positions of
the QR.

\begin{figure}[!t]
  \centering
  \includegraphics[width=8cm]{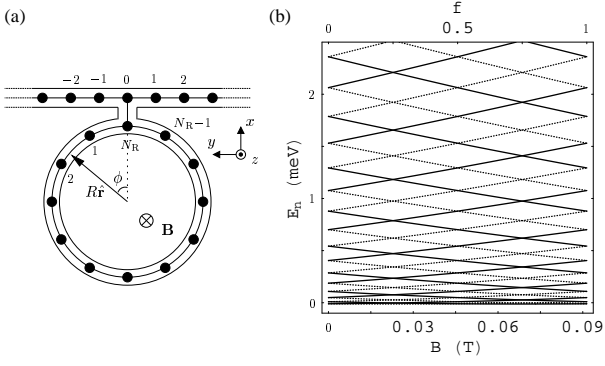}
  \caption{(a) Schematic view of a quantum ring side-coupled to a
    quantum wire. Both systems are described by tight-binding models.
    The number of sites in the quantum ring is $N_{\rm R}$; in our
    study the limit $N_{\rm R}\to\infty$ is taken. (b) Energy spectrum
    of a non-interacting QR as a function of the magnetic field $B$ or
    the corresponding flux $f$ threading the QR with a radius
    $R=120\nmeter$. The dotted and solid lines correspond to spin index
    $\mu=+$ and $-$ levels, respectively. Here, the material
    parameters for GaAs are taken such that $m = 0.067m_e$ and $\alpha
    = 0.53\times10^{-11} \evolt\meter$ in which case $2/|\cos\theta| =
    3$.}
  \label{fig:1}
\end{figure}

In this Letter we propose another kind of spin filter that consists of
a QR \textit{side-coupled to} a QW; see
\figref{fig:1}(a). Spin-resolved Fano resonances due to spin-split
levels formed in the QR in the presence of Rashba spin-orbit coupling
and external magnetic fields \cite{SOQuantumRing} lead to a complete
suppression of transport of either spin component at a set of gate
voltage values, resulting in a series of valleys in the spin-resolved
conductance.  The separations between valleys are observed to be of
the order of the Coulomb interaction energy.  This QR-based spin
filter has three advantages: (1) It does not require strong magnetic
fields and high $g$-factors like the QD-based system.  In the presence
of spin-orbit coupling, a weak magnetic field applied to a small QR
\cite{QuantumRing} can induce a large energy splitting between spin
levels because it is the magnetic flux
that causes the level splitting in the ring geometry.  (2) Only a single
contact of the QR to the external circuit such as leads or wires is necessary.
(3) Finally, on-board tuning of polarization direction of spin currents is
possible via the control of spin-orbit coupling strength.

\textit{Model.}--- First, we examine the energy level structure of a
non-interacting QR with Rashba spin-orbit coupling in the presence of
an external magnetic field. In the ideal one-dimensional limit where the radial
width is much smaller than the radius $R$, only the lowest
radial subband is occupied \cite{Ideal1DRing}, and the effective
Hamiltonian projected to the lowest radial mode can be written in
polar coordinates \cite{SOQuantumRing}
\begin{align}
  \varH_{\rm RN} & = \frac{\hbar^2}{2mR^2}
  \left(i\Pde{\phi} + f\right)^2
  - \frac{\alpha}{2R} \{\bfsigma\cdot\ur,i\Pde{\phi} + f\},
\end{align}
where $m$ is the effective electron mass, $\bfsigma$ the Pauli
matrices, $\ur$ a unit vector in radial direction, and $\{A,B\} = AB + BA$.
The energy of the radial mode is omitted.
The spin-orbit coupling strength
$\alpha$ defines the spin-flip length $l_{\rm so} \equiv
\pi\hbar^2/m\alpha$, and the external magnetic field induces the
normalized magnetic flux $f \equiv \pi BR^2/\Phi_0$ threading the
ring; $\Phi_0 = hc/e$ is the flux quantum. The Zeeman splitting can be
ignored compared to the kinetic energy $E_0\equiv \hbar^2/2mR^2$ as
long as $1 \gg g\mu_B B/E_0 = g (m/m_e) f$ is well satisfied;
this holds in usual semiconducting materials with $f \lesssim 1$.
Following the standard procedure \cite{Datta95}, one can set up the
tight-binding version of the Hamiltonian in terms of spin-$\mu$ electron
operators $a_{n\mu}$, $a^\dagger_{n\mu}$ defined at site $n$ of the
QR, and the tight-binding Hamiltonian can be diagonalized through the
Fourier transformation such that
\begin{align}
  \varH_{\rm RN} & = \sum_{m=1}^{N_{\rm R}} \sum_{\mu=\pm}
\epsilon_{m\mu} d_{m\mu}^\dag d_{m\mu}.
\end{align}
The operator $d_{m\mu}^\dag$ creates an electron in orbital mode $m$
and with spin index $\mu$ having space-dependent polarization $\hat\bfn = \uz
\cos\theta + \ur \sin\theta$ that does not depend on the orbital index $m$
due to the assumption that the Zeeman splitting
is negligible \cite{SOQuantumRing}.  In the limit $N_{\rm R}\to\infty$, the eigenenergy
$\epsilon_{m\mu}$ is given by
\begin{align}
  \label{eq:specone}
  \epsilon_{m\mu} & = E_{0}\!
  \left[\left(m {+} f {-} \frac12 {+} \frac{\mu}{2\cos\theta}\right)^2
    + \frac14 \left(1 {-} \rcp{\cos^2\theta}\right)\right]
\end{align}
with the polarization angle $\theta=\arctan\left[-N_{\rm so}\right]$,
where $N_{\rm so}\equiv 2\pi R/l_{\rm so}$ is the number of spin flips
around the ring. The resulting energy spectrum is periodic not only in
$f$ but also in $1/2\cos\theta$ (excluding the overall shift due to the
last term in \eqnref{eq:specone}).  Moreover, the energy gaps
between neighboring spin-split levels reach their maxima whenever
$1/2|\cos\theta| = (2l+1)/4$, whereas the spin-splitting disappears at
$1/2|\cos\theta| = l/2$, for integer $l>1$. Throughout our Letter
$2/|\cos\theta|$ is assumed to be an odd integer to maximize
the spin-splitting.
\Figref{fig:1}(b) shows the energy spectrum for realistic material
parameters for GaAs. The ring size is taken to be $R = 120\nmeter$,
which is feasible using current fabrication technology
\cite{QuantumRing}. The spectrum shows that a small magnetic field $<
50\mtesla$ is enough to induce a spin-splitting energy gap comparable
to $1$ to $3\kelvin$. This large splitting that exists even in the absence of a
strong external magnetic field definitely makes the QR a good
candidate for ideal spin filter operation.

To take into account the electron-electron Coulomb
interaction in the small QR, we adopt a simple capacitive model where
the Coulomb interaction depends only on the total number of electrons:
$\varH_{\rm RI} = (U/2) \left[N^2 - 2N_g N\right]$ with $N {\equiv}
\sum_{m\mu} d_{m\mu}^\dag d_{m\mu}$.  Here $U\equiv e^2/(C+C_g)$ and
$N_{\rm g}\equiv C_gV/|e|$ denote the interaction strength and the gate charge
(in units of $|e|$), respectively, in terms of self and gate capacitances, $C$ and $C_g$.

The total Hamiltonian for a QR side-coupled to a QW can then be
written as
\begin{align}
  \varH = \varH_{\rm RN} + \varH_{\rm RI} + \varH_{\rm W} + \varH_{\rm WR}
\end{align}
with $\varH_{\rm W} = - t_{\rm w} \sum_{n\mu} (c_{n+1\mu}^\dag
c_{n\mu} + h.c.)$ and $\varH_{\rm WR} = t_{\rm wr} \sum_{\mu}
(c_{0\mu}^\dag a_{N_{\rm R}\mu} + h.c.)$, where the operators $c_{n\mu}$
($c_{n\mu}^\dag$) destroys (creates) an electron with spin index $\mu$
at site $n$ of the wire. $\varH_{\rm W}$ models the QW as an infinite
tight-binding chain with a hopping energy $t_{\rm w}$ between
neighboring sites, and $\varH_{\rm WR}$ a spin-independent tunneling
with strength $t_{\rm wr}$ between site 0 of the wire and site $N_{\rm R}$
of the ring. Note that the spin quantization axis for the QW has been
rotated to align with the spin axis at site $N_{\rm R}$ of the QR,
$\hat\bfn$ at $\ur=\ux$.

\begin{figure}[!t]
  \centering
  \includegraphics[width=8.5cm]{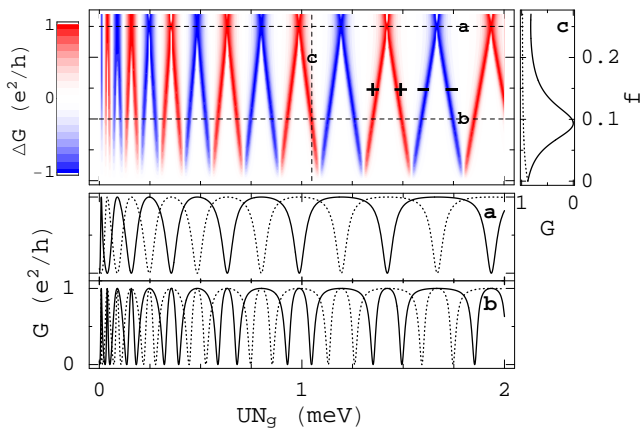}
  \caption{(Color online) Net spin conductance $\Delta G$ as a function of
    gate voltage $UN_g$ and magnetic flux $f$ in the non-interacting case
    at zero temperature. The plus and
    minus signs indicate the sign of $\Delta G$ and are assumed to be
    repeated periodically along the $UN_g$ axis. Here we have used the same
    QR parameter values as in \figref{fig:1}(b) and set $t_{\rm w} = 5\mevolt$,
    $t_{\rm wr} = 0.4\mevolt$. The right and bottom figures show the spin-resolved conductances
    $G_+$ (dotted) and $G_-$ (solid) taken along the dashed lines in the main figure.}
  \label{fig:2}
\end{figure}

\textit{Spin filter.}--- We have calculated the zero-bias conductance
$G_\mu$ for spin $\mu$ at the Fermi level $\epsilon_F = 0$ under the
assumption that two electron reservoirs with nearly the same chemical
potentials are attached at both ends of the QW \cite{EntinWohlman86}.
The non-equilibrium scattering formalism \cite{Meir92} enables us to
express the conductance in terms of the Green's function
$\varG^R_\mu(\epsilon)$ for a spin-$\mu$ electron at site 0 of the
QW:
\begin{align}
  \label{eq:G}
  G_\mu & = \frac{e^2}{h} \int d\epsilon \pde{f(\epsilon)}{\epsilon}
\im \Gamma(\epsilon) \varG^R_\mu(\epsilon)\; .
\end{align}
Here, $f(\epsilon)$ is the Fermi distribution function with
$\epsilon_F = 0$ and the symmetric coupling $\Gamma(\epsilon)$ of site
0 to the left and right sides of the QW is given by $\Gamma(\epsilon)
= (2t_{\rm w}/\hbar) \sin\chi_{\rm w}(\epsilon)$ with $\chi_{\rm
w}(\epsilon) \equiv \arccos [- \epsilon/2t_{\rm w}]$. In the
non-interacting case ($C\ll C_g$ and $U\approx 0$), by solving the Dyson equation for
$\varG^R_\mu$, we obtain the spin-dependent transmission probability
$T_\mu(\epsilon) = -\im \Gamma(\epsilon) \varG^R_\mu(\epsilon) = 1/(1
+ [Q_\mu(\epsilon)]^2)$ with $Q_\mu(\epsilon) \equiv
\Gamma(\epsilon)^{-1} (t_{\rm wr}/\hbar)^2 \sum_m g^R_{m\mu}(\epsilon)
= \Delta(\epsilon) \sum_m 1/(\epsilon - \epsilon_{m\mu})$, where
$g^R_{m\mu}$ is the Green's function for the uncoupled QR and
$\Delta(\epsilon)\equiv t_{\rm wr}^2/\hbar\Gamma(\epsilon)$ is the
level broadening due to the QW-QR coupling.  Since $Q_\mu$ diverges at
$\epsilon = \epsilon_{m\mu}$, the transmission probability $T_\mu$
vanishes whenever a resonant state with spin $\mu$ is formed in the
QR, giving rise to perfect suppression of the transport of spin-$\mu$
electrons.  Note that this blocking condition is independent of any
characteristics of the wire.  \Figref{fig:2} shows the formation of a
series of spin-split dips in the zero-bias conductances $G_\mu$ as
functions of the gate voltage at zero temperature.
The width of the valleys is restricted by the minimum of
the energy splitting between neighboring levels and the level
broadening $\Delta(\epsilon_F)$. The spin-dependent conductance can
also be controlled by varying the flux $f$. The condition for total transmission
($T_\mu=1$), $\sin 2\pi\sqrt{(\epsilon_F-UN_g)/E_0} = 0$, does
not depend on spin, thus the peak positions in $G_\mu$ are the same
for both spins.

This spin-dependent transmission generates a net spin flow through the
wire: $\Delta G \equiv G_+ - G_- = (e^2/h) ([Q_-]^2 - [Q_+]^2)/[(1 +
[Q_+]^2)(1 + [Q_-]^2)]$ at zero temperature.  The net spin conductance
$\Delta G$ has local maxima or minima whenever one of the $Q_\mu$
diverges; see \figref{fig:2}. The peak height in $|\Delta G|$ reaches
almost the maximum value $e^2/h$ if the spin splitting
$\delta\epsilon$ is larger than the broadening $\Delta(\epsilon_F)$,
in which case the unblocked states with opposite spin are transmitted
almost copmletely. The ideal operation of the spin filter, therefore, requires
$\delta\epsilon\gg\Delta(\epsilon_F)$. In addition, to avoid
temperature-induced broadening through \eqnref{eq:G}, both the spin
splitting and the broadening should be larger than the temperature $T$
as well.
Interestingly, at $f=1/4$, $|\Delta G|$ reaches $e^2/h$ at its peaks
regardless of $UN_g$, implying perfect blocking for one spin direction and perfect transmission
for the other. Also, the peak widths are maximal  at $f=1/4$. This is
related to the appearance of degenerate levels with the same spin at
$f=1/4$ [see \figref{fig:1}(b)], which merges two peaks separated at
$f\ne1/4$ into one peak and strengthens the Fano resonances with
broader width.  This observation indicates that the best performance
of the spin filter can be achieved at $f=1/4$.

It should be noted that the polarization direction
$\hat\bfn=\uz\cos\theta+\ux\sin\theta$ of the spin current can be rotated
by tuning the strength of the spin-orbit coupling, which should be still
adjusted to satisfy the odd-integer condition of $2/|\cos\theta|$ to
achieve the maximal separation between spin-split levels.  If the spin
current is measured along a direction $\hat\bfn'$ other than
$\hat\bfn$, the net spin current decreases via $\Delta G|_{\hat\bfn'}
= \Delta G|_{\hat\bfn} \cos\zeta$, where $\zeta$ is the relative angle
between $\hat\bfn$ and $\hat\bfn'$.

\textit{Coulomb interaction.}--- We now turn on the self-charging interaction
in the QR with moderate values of $U$ and investigate its effect on
the transport at finite temperatures. The numerical renormalization
group method, proven to be an excellent numerical tool for
Anderson-type impurity systems, was applied to calculate the
spin-resolved local density of states $\rho_\mu(\epsilon)$ on site
$N_R$ of the ring.  The transmission amplitude can then be calculated
using the Dyson equation: $T_\mu(\epsilon) = 1 - \pi \Delta(\epsilon)
\rho_\mu(\epsilon)$.

\begin{figure}[!t]
  \centering
  \includegraphics[width=7cm]{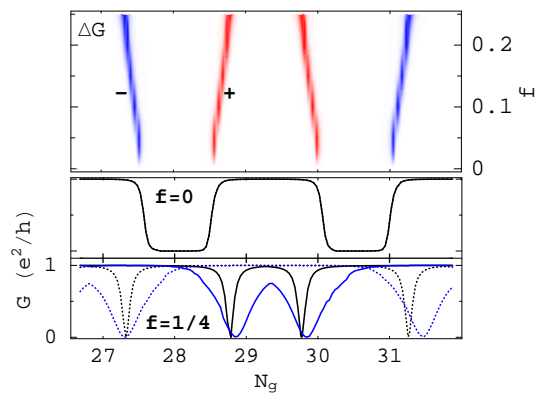}
  \caption{(Color online) Net spin conductance $\Delta G$ as a function of
    gate charge $N_g$ and magnetic flux $f$ in the interacting case with $U = 0.5\mevolt$.
    Lower figures show spin-resolved conductances $G_\pm$ for $f=0$ and $1/4$, respectively.
    The same plot styles and parameter values as in \figref{fig:2} are used.
    For $f=1/4$, the conductances (blue lines)
    with a larger coupling $t_{\rm wr} = 0.8\mevolt$ are also shown.}
  \label{fig:3}
\end{figure}

\Figref{fig:3} shows the dependence of the zero-temperature
conductances $G_\mu$ and $\Delta G$ on magnetic flux and gate voltage
which has been tuned to such
large values that high QR levels with $\delta\epsilon \gg
\Delta(\epsilon_F)$ contribute to the transport: the dips in
conductance correspond to the QR levels with $m=13$ and $14$.
At $f=0$ the correlation between spin-degenerate QR levels and QW
conduction electrons induces the Kondo effect whenever the QR contains
an odd number of electrons. As a consequence, the Fano resonance due
to the resultant effective resonant level at the Fermi level
suppresses the charge transport regardless of the spin direction
\cite{Kang01}. Each broad valley in $G_\mu$ at $f=0$, however, splits
into two spin-dependent sharp dips as the spin-splitting
$\delta\epsilon$ due to finite magnetic field $(f\ne0)$ becomes larger
than the Kondo temperature $T_K$.
The Coulomb repulsion widens the separations between dips in $G_\mu$ or $\Delta G$,
which is now of the order of $U$ \cite{Torio04,Aligia04}, or
$\varO(1)$ in terms of $N_g$ even at $f=1/4$, while
the width of valleys, still of the order of $\Delta(\epsilon_F)$, is
not affected. As long as $U>\Delta(\epsilon_F)$, the broadened
separation due to the Coulomb interaction contributes toward perfect
spin filtering at $f\ne1/4$. As in the non-interacting case, the dip width
increases as $f$ goes to 1/4.
For large QW-QR coupling, but still
$\delta\epsilon>\Delta(\epsilon_F)$, the valleys can overlap, opening
wide gate-voltage windows for inducing a finite spin current; see the
uppermost figure in \figref{fig:3}. Larger coupling with
$\delta\epsilon\le\Delta(\epsilon_F)$, however, leads to concurrent
suppression of both spins and smaller net spin current.

\begin{figure}[!t]
  \centering
  \includegraphics[width=7cm]{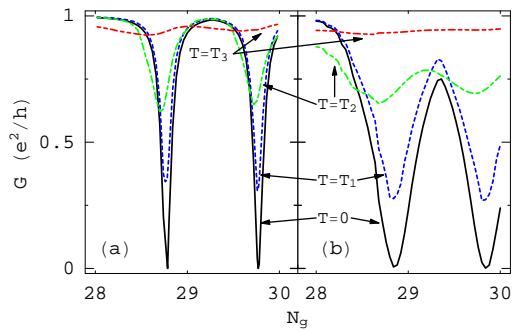}
  \caption{(Color online) Finite-temperature conductance $G_-$ for
    $f=1/4$. The temperatures are given by $k_BT_1 =
    0.13\Delta(\epsilon_F)$, $k_BT_2 = \Delta(\epsilon_F)$, and
    $k_BT_3 = 8.1\Delta(\epsilon_F)$.
    (a) $t_{\rm wr} =0.4\mevolt$  and $\Delta(\epsilon_F)=0.016\mevolt$.
    (b) $t_{\rm wr} =0.8\mevolt$ and $\Delta(\epsilon_F)=0.064\mevolt$.
    Other parameters as in \figref{fig:3}.}
  \label{fig:4}
\end{figure}

Thermal fluctuations diminish the resonance-based suppression as soon
as $k_BT\gtrsim\Delta(\epsilon_F)$; see \figref{fig:4}.
First, thermal broadening in the QW, via the smoothened peak in
$\partial f/\partial\epsilon$, obscures the resonance as the
temperature becomes comparable to the resonance width in
$\rho_\mu(\epsilon)$, or the broadening $\Delta(\epsilon_F)$.
Second, thermal fluctuations invoke transitions between QR levels that
also weaken the resonance and consequently diminish the peak height
in $\rho_\mu(\epsilon)$.  The rapid degradation of the spin filter effect
at $k_BT \approx \Delta(\epsilon_F)$ can be attributed to these
thermal fluctuations.
Consequently, our system will show ideal spin-filter operation if
$k_BT\ll\Delta(\epsilon_F)\ll\delta\epsilon$.  Since
$k_BT_K\ll\delta\epsilon $ in most cases with $f\ge0.1$, the Kondo
temperature is irrelevant in spin filtering.

\textit{Discussion.}--- If the ring width is not narrow compared to
the radius, higher radial modes will contribute to the transport.
While the large energy gap between radial modes prohibits the direct
excitation to higher modes at low temperatures, the spin can
experience dephasing or relaxation due to the spin-orbit interaction
that couples different radial modes with opposite spins
\cite{SOQuantumRing}. Also, nonmagnetic impurity scattering in such a thick ring
with spin-orbit coupling can smear out the spin filtering.

\textit{Conclusions.}--- Our spin filter takes advantage of two
ingredients: (1) the relatively large spin-splitting in a small QR with
Rashba spin-orbit coupling and (2) the Fano resonances due to the
spin-split levels in the QR that is side-coupled to a QW with one
conduction channel.  We predict perfect or considerable suppression of
the transport of either of spin direction under real experimental
conditions that are accessible using current technology.

\begin{acknowledgments}
  We would like to thank V. Golovach and M.-S. Choi for helpful
  discussions.  This work was financially supported by the SKORE-A
  program, the Swiss NSF, and the NCCR Nanoscience.
\end{acknowledgments}

\end{document}